\documentclass[aps,prd,twocolumn,superscriptaddress,showpacs,nofootinbib,eqsecnum,amsfonts,amsmath,groupedaddress]{revtex4-1}

\usepackage{amsfonts}
\usepackage{amsmath}
\usepackage{amssymb}
\usepackage{slashed}
\usepackage{graphicx,color}
\usepackage{float}
\usepackage{hyperref}

\def\nn{\nonumber}
\def\be{\begin{equation}}
\def\ee{\end{equation}}
\def\bea{\begin{eqnarray}}
\def\eea{\end{eqnarray}}
\def\ms{\overline {\rm MS}}

\begin{document}

\title{Scale independence in an asymptotically free theory at finite temperatures} 

\author{Gabriel N. Ferrari}
\email{gabrielferrari@univali.br} 
\affiliation{Universidade do Vale do Itaja\'{\i}, 88302-901 Itaja\'{\i}, Santa Catarina, Brazil\linebreak Departamento de F\'{\i}sica, Universidade Federal de Santa
  Catarina, 88040-900 Florian\'{o}polis, Santa Catarina, Brazil}

\begin{abstract}

 A recently developed variational resummation technique incorporating renormalization group
properties has been shown to solve the scale dependence problem that plagues the
evaluation of thermodynamical quantities, e.g., within the framework of approximations such as in
the hard-thermal-loop resummed perturbation theory. This method is used in the present work
to evaluate thermodynamical quantities within the two-dimensional nonlinear sigma model, which shares some features with Yang-Mills theories like asymptotic freedom, trace anomaly and the nonperturbative generation of a mass gap. Besides the fact that nonperturbative results can be readily generated solely by
considering the lowest-order contribution to the thermodynamic effective potential, we also show that its next-to-leading correction indicates convergence to the sought-after scale
invariance.

\end{abstract}

\maketitle

\section{Introduction}

The theoretical description of the quark-gluon plasma
phase transition requires the use of nonperturbative
methods, since the use of perturbation theory near
the transition is unreliable. LQCD has been very successful at finite temperatures and near
vanishing baryonic densities, however, currently, the complete description of compressed
baryonic matter cannot be achieved due to the
so-called sign problem. In this case, an alternative is
to use approximate but more analytical nonperturbative
approaches. One of these is the Optimized Perturbation Theory (OPT), which reorganizes the series using
a variational approximation, where the result of a related
solvable case is rewritten in terms of a variational parameter that allows for nonperturbative results to be obtained. On the other hand, the results of the Hard
Thermal Loop Perturbation Theory (HTLpt), done
in a gauge-invariant framework, exhibit a strong sensitivity to the arbitrary
renormalization scale M used in the regularization
procedure~\cite{HTLPT3loop}. A solution to this problem has been recently proposed,
by generalizing to thermal theories a related variational
resummation approach, Renormalization Group Optimized
Perturbation Theory (RGOPT)~\cite{JLGN}. In this work we apply the RGOPT to the nonlinear
sigma model (NLSM) in 1+1 dimensions at finite temperatures
in order to pave the way for future applications
concerning other asymptotically free theories, such
as thermal QCD.  As we will illustrate, the scale invariant results obtained
in the present application give further support to
the method as a robust analytical nonperturbative approach
to thermal theories.

\section{The NLSM in 1+1-dimensions}
\label{sec2}

The two-dimensional
NLSM partition function can be written as~\cite{nlsmrenorm}

\begin{equation}
Z \!=\! \int \! \prod_{i=1}^{N} {\cal D} \Phi_i(x) \exp \left [  \frac
  {1}{2 g_0} \!\!\int \! d^2 x (\partial \Phi_i)^2 \right ] \delta
(\sum_{i=1}^N \Phi_i \Phi_i -1),
\label{action}
\end{equation}
where $g_0$ is a (dimensionless) coupling and the scalar field is parametrized as
$\Phi_i=(\sigma,\pi_1,...,\pi_{N-1})$. In two-dimensions the theory is
 renormalizable~\cite{nlsmrenorm}. The action is invariant under
$O(N)$ but using the constraint, $\sigma(x)= (1-  \pi_i^2)^{1/2}$, 
 in order to define the perturbative expansion, 
breaks the symmetry down to
$O(N-1)$. In this case the partition function becomes 
\begin{equation}
Z(m) = \int d\pi_i(x)  \left [1- \pi_i^2(x) \right ]^{-1/2}  \exp
[-{\cal S}(\pi,m)],
\end{equation}
where the (Euclidean) action is ${\cal S}(\pi,m) = \int d^2 x {\cal
  L}_0$ and, upon rescaling $\pi_i \to\sqrt{ g_0} \pi_i$, we can read the  bare
Lagrangian density and expand it to order-$g_0$ yielding
\begin{equation}
{\cal L}_0 \!=\! \frac{1}{2}  \left[ (\partial \pi_i)^2 + m_0^2
  \pi_i^2 \right] + \frac {g_0 m_0^2}{8} (\pi_i^2)^2 + \frac{g_0}{2 }
(\pi_i \partial \pi_i)^2 -{\cal E}_0 ,
\end{equation}
 where for later notational convenience we designate as ${\cal E}_0\equiv m^2_0/g_0$ 
the field-independent term, originating at lowest order from expanding the square root in the bare lagrangian.

In this work,  the divergent integrals are regularized
using dimensional regularization (within the minimal subtraction
scheme $\overline {\rm MS}$), which at finite temperature and $d=2-
\epsilon$ dimensions, can be implemented by using
\begin{equation}
\int \frac {d^2 p} {(2\pi)^2} \to T \,
\hbox{$\sum$}\!\!\!\!\!\!\!\int_{\bf p} \equiv T\left ( \frac
     {e^{\gamma_E} M^2}{4\pi} \right)^{\epsilon/2}
     \sum_{n=-\infty}^{+\infty} \int \frac {d^{1-\epsilon} 
         p}{(2\pi)^{1-\epsilon}},
\end{equation}
 where $\gamma_E$ is the Euler-Mascheroni constant and $M$ is the
 $\overline {\rm MS}$ arbitrary regularization energy scale.  

\section{ Perturbative Pressure and Scale Invariance}\label{sec3}
 \begin{figure}[htb!]
\includegraphics[scale=0.5]{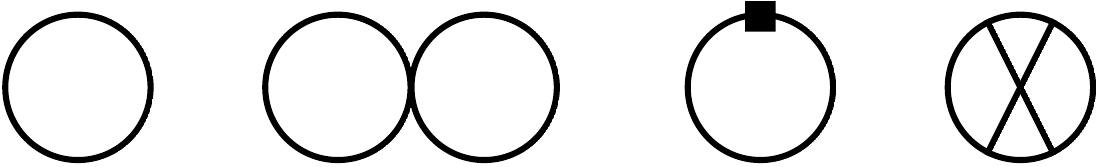}
\caption[long]{\label{fig1} Feynman diagrams contributing to the
  perturbative pressure at ${\cal O}(g)$. The first term represents $P_0(m_0)$, the second, $P_1(m_0, g_0)$, 
  the third term represents the self-energy counterterm $P_0^{\rm CT}$ (obtained 
  from expanding $Z_m$ to first order in $P_0(m_0=Z_m\,m)$), while the fourth term
  represents the zero point contribution ${\cal E}_0(g_0)$ to
  Eq.~(\ref {press2l}). }
\end{figure}
Considering the  contributions displayed in
{}Fig.~\ref{fig1}, one can write the pressure up to order ${\cal O}(g_0)$ as
\begin{equation}
 P= P_0(m_0) + P_1(m_0, g_0) + {\cal E}_0(m_0, g_0) + {\cal O}(g_0^2).
\label {press2l}
\end{equation}

By implementing renormalization consistently after the identification of the counterterms (details can be found on \cite{ournlsm}), the renormalized two-loop pressure can be written in its compact form
\begin{equation}
P = \frac{m^2}{g} - \frac{(N-1)}{2} \left[ I_0^{\rm r}(m,T) + \frac{(N-3)}{4} m^2 g
\left[I_1^{\rm r} (m,T)\right]^2 \right],
\label{RenP}
\end{equation}  
where
\begin{eqnarray}
I_0(m_0,T) =  T \, \hbox{$\sum$}\!\!\!\!\!\!\!\int_{\bf p} \ln \left(
\omega_n^2+\omega_{\bf p}^2 \right),
\end{eqnarray}
with the dispersion relation,  $\omega_{\bf p}^2 ={\bf p}^2 +m_0^2$ and $I_1(m_0,T)=\partial I_0(m_0,T)/\partial m_0^2$.

Considering the renormalization group (RG) operator,  defined by
\begin{equation}
M \frac{d}{dM} \equiv  M \frac{\partial}{\partial M} + \beta
\frac{\partial}{\partial g} - \frac{n}{2} \zeta + \gamma_m m
\frac{\partial}{\partial m} .
\label{RG}
\end{equation} 
Applying the latter to the pressure (zero-point vacuum energy) one has $n=0$,
so that one only needs to consider the $\beta$ and $\gamma_m$ functions.
At the two-loop level,
\begin{equation}
\beta = - b_0 g^2 - b_1 g^3 + {\cal O}(g^4) ,
\label{beta}
\end{equation}
and
\begin{equation}
\gamma_m = -\gamma_0 g-\gamma_1 g^2 +{\cal O}(g^3),
\label{gammam}
\end{equation}
where the RG coefficients in our normalization are~\cite{hikami}:
\begin{eqnarray}
&& b_0=(N-2)/(2\pi),
\label{b0}
\\ && b_1=(N-2)/(2\pi)^2,
\label{b1}
\\  &&\gamma_0=(N-3)/(8\pi),
\label{gamma0}
\\ &&\gamma_1=(N-2)/(8\pi^2).
\label{gamma1}
\end{eqnarray}

Following \cite {jlprl} one can write the finite zero-point
  energy contribution, ${\cal E}_0^{\rm RG}$: 
\begin{equation}
{\cal E}_0^{\rm RG} = m^2 \sum_{k \ge 0} s_k g^{k-1}  ,
\label{skdef}
\end{equation}
and determine the coefficients $s_k$  by applying (\ref{RG}) consistently order by order.
In the present NLSM, one can easily check that it uniquely fixes the relevant coefficients
up to two-loop order, $s_0, s_1$, as
\begin{equation}
s_0 = \frac{(N-1)}{4\pi (b_0-2 \gamma_0)} = 1 ,
\label{s0}
\end{equation}
and
\begin{equation}
s_1= (b_1-2\gamma_1)\frac{s_0}{2 \gamma_0}=0,
\label{s1}
\end{equation}
(which vanishes as $b_1=2\gamma_1$ in the NLSM). \\
Thus from perturbative RG considerations, Eq.~(\ref{skdef}) with (\ref{s0}), (\ref{s1}) 
reconstructs consistently the NLSM first term of (\ref{RenP}),
originally present in our original NLSM derivation above. RG invariance is maintained 
(or more correctly, restored) also
within the more drastic modifications implied by the variationally optimized perturbation framework, as we examine now. 

\section{RG Improved Optimized Perturbation Theory}
\label{sec4}
 
 To implement next the RGOPT one first modify the standard perturbative
 expansion by rescaling the infrared regulator $m$ and coupling:   \be
 m\to (1-\delta)^a m,\;\;\;\; g \to \delta g,
 \label{delta}
 \ee in such a way that the Lagrangian interpolates between a free
 massive theory (for $\delta=0$) and the original massless theory
 (for $\delta=1$)~\cite{JLQCD1}. Since the mass parameter is being optimized by using the
variational stationary mass optimization
prescription~\cite{pms}, as in OPT or in the Screened Perturbation Theory (SPT, ~\cite{spt}),
\begin{equation}
\frac{\partial P^{\rm RGOPT}}{\partial m} \Bigr|_{m={\bar m}} = 0 ,
\label{pms}
\end{equation}
the RG operator acquires the {\it reduced} form

\begin{equation}
\left( M \frac{\partial}{\partial M} + \beta \frac{\partial}{\partial
  g}  \right) P^{\rm RGOPT}=0 .
\label{RGr} 
\end{equation}
which is indeed consistent for a massless theory.

Then, performing the aforementioned replacements given by
Eq.~(\ref{delta}) within the pressure Eq.~(\ref{RenP}), consistently re-expanding to lowest (zeroth) order in
$\delta$, and finally taking $\delta\to 1$,  one gets
\begin{equation}
P^{\rm RGOPT}_{ 1L} =  - \frac{(N-1)}{2} I_0^{\rm r}(m,T)+
\frac{m^2}{g} (1 - 2a).
\label{Prgopt}
\end{equation}
Now to fix the exponent $a$ we require the RGOPT pressure, Eq.~(\ref
{Prgopt}), to satisfy the {\it reduced} RG relation, Eq.~(\ref
{RGr}). This {\it uniquely} fixes the exponent to 

\begin{equation}
a = \frac{\gamma_0}{ b_0} =\frac{(N-3)}{4(N-2)}.
\label{exponent}
\end{equation}

With the exponent $a$ determined, one can write the resulting one-loop
RGOPT expression for the NLSM pressure as

\begin{equation}
P^{\rm RGOPT}_{ 1L} =  - \frac{(N-1)}{2} I_0^{\rm r}(m,T)+ (N-1)
\frac{m^ 2}{(4\pi) g b_0}.
\label{P1L}
\end{equation}
In the same way, the two-loop standard PT result obtained in the
previous section gets modified accordingly to yield the corresponding
RGOPT pressure at the next order of those approximation
sequences. After performing the  substitutions given by
Eq.~(\ref{delta}),  with $a=\gamma_0/b_0$ within the two-loop PT
pressure Eq.~(\ref{RenP}),   expanding now to first order in $\delta$,
next taking the limit $\delta\to 1$, gives
\begin{eqnarray} 
P^{\rm RGOPT}_{ 2L} &=& -\frac{(N-1)}{2} I_0^{\rm r}(m,T) +  (N-1)
\left(\frac{\gamma_0}{b_0}\right ) m^2 I_1^{\rm r}(m,T)\nonumber \\ &
-&  g (N-1) \frac{(N-3)}{8}  m^2 \left[I_1^{\rm r}(m,T)\right]^2 \nonumber \\ &+&
\frac{(N-1)}{4\pi} \frac{   m^2}{g \, b_0 }\left (1-
\frac{\gamma_0}{b_0}\right ) .
\label{rgopt2L}
\end{eqnarray}

\subsection*{RGOPT mass gap and running coupling constant}

By optimizing the pressures above and solving the mass gap \cite{ournlsm}, at one-loop one obtains 

\begin{equation}
{\bar m}(0) = M \exp\left( -\frac{1}{b_0\,g(M)} \right).
\label{mass1L}
\end{equation}
 It is instructive to remark that the above optimized mass
 gap is dynamically generated by the (nonlinear) interactions and
 reflects dimensional transmutation, with nonperturbative coupling dependence. 

The Eq.~(\ref{mass1L}) moreover fixes the optimized mass $\bar m$ to be fully
 consistent with the running coupling $g(M)$ as described by the usual
 one-loop result,

\begin{equation}
g^{-1}(M) = g^{-1}(M_0) + b_0 \ln\frac{M}{M_0} ,
\label{run1}
\end{equation}
where $M_0$ is an arbitrary reference scale and $L = \ln (M/M_0)$.  

Going now to two-loop order, the mass optimization criterion
Eq.~(\ref{pms}) applied to the RGOPT-modified two-loop pressure
Eq.~(\ref{rgopt2L}) can be cast, after straightforward
algebra, in the form (omitting some irrelevant overall factors):
\bea 
& f_{MOP}^{(2L)}\equiv \left\{ 3N-5 -b_0 (N-3)\, g \,\left[1 +Y
  +2x^2 J_2(x)\right]\right\}   \nn \\ &\times m \,\left(\frac{1}{b_0\,g}
+Y \right)   = 0,
\label{pms2loop}
\eea   
where, we have defined for convenience the following dimensionless quantity
\be 
Y \equiv \ln \frac{m}{M}+2J_1(x) = -2\pi\,I_1^r(m,T),
\label{Ydef}
\ee 
and the thermal integral, $x J_2(x)\equiv \partial_x J_1(x)$ reads
\begin{equation}
J_2(x) = \int_0^\infty dz
\frac{[e^{\omega_z}(1+\omega_z)-1]}{\omega_z^3 (1-e^{\omega_z})^2} .
\end{equation}
Alternatively, the reduced RG equation (\ref{RGr}), using the exact
two-loop $\beta$-function Eq.~(\ref{beta}), yields
\bea 
& f_{RG}^{(2L)} \equiv  m^2\,\left[ g \frac{(3N-5)}{2\pi} +(N-3) \times \right. \nn \\ 
& \left. \left\{ 1+\frac{N-2}{\pi} g \,Y \left[1+ \frac{N-2}{4\pi} g\,
  (1+\frac{g}{2\pi})\,Y\right]\right\} \right]=0 .
\label{rg2loop}
\eea 

When considered as two alternative (separate) equations, 
(\ref{pms2loop}) and (\ref{rg2loop}), apart from having the trivial solution $\bar m=0$, 
also have a more interesting nonzero mass gap solution, $\bar
m(g,T,M)$, with nonperturbative dependence on the
coupling $g$. 

Hence, apart from the one-loop running in Eq.~(\ref{run1}), we
also need the two-loop running coupling, with exact expression given e.g. in ~\cite{jlprd}, which can be
approximated as follows with sufficient accuracy (as long as $g$
remains rather moderate $g\sim {\cal O}(1)$),
\begin{eqnarray} 
g^{-1}(M) &\simeq & g^{-1}(M_0) +b_0 L   +(b_1 L)g(M_0) \nonumber
\\  &-&\left (\frac{1}{2} b_0 b_1 L^2\right )  g^2(M_0) \nonumber
\\  & -&\left (\frac{1}{2} b_1^2 L^2 -\frac{1}{3} b_0^2 b_1 L^3\right
) g^3(M_0)  \nonumber \\ &+& {\cal O}(g^4) .
\label{run2}
\end{eqnarray}

\section{Numerical Results}
\label{sec5}

To investigate and compare  the scale variation behavior of the
different approximations - such as large-$N$ (LN) and SPT -  in our analysis below, as it is customary,
we set the arbitrary $\ms$ scale  as $M=\alpha M_0=2\pi T\alpha $ and
consider $0.5 \leq \alpha \leq 2$ as representative values of scale
variations. On \cite{ournlsm} one can find the explanation of the residual scale dependence and a complete description of the parameters choice regarding the coupling.
%

\begin{figure}[htb!]
\includegraphics[scale=0.95]{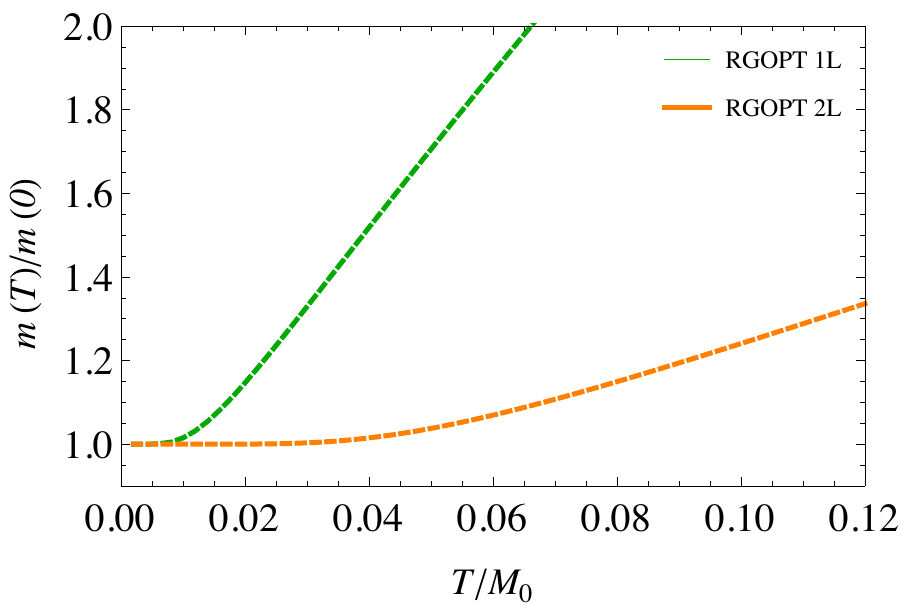}
\caption{\label{fig4} The normalized thermal optimized masses, ${\bar
    m}(T)/{\bar m}(0)$, as a function of the temperature $T$
  (normalized by $M_0$)   for $N=4$, $g(M_0)=1 = g_{LN}(M_0)/2$, and
  at the central scale choice $\alpha=1$, in the RGOPT  at one- and
  two-loop cases.  (NB for this coupling choice the LN thermal mass is
  identical to the RGOPT one-loop one).}
\end{figure}

The RGOPT mass ${\bar m}(T)$  clearly starts from a nonzero value at
$T=0$, since the mass gap solution is nontrivial at $T=0$ (Eq.~(\ref{mass1L})), then bends
and reaches, as expected by using basic dimensional arguments, a
straight line for large temperatures, where it behaves perturbatively as $\bar m\sim g T$.  As observed in \cite{giacosa},  this behavior is reminiscent of that of the
gluon mass in the deconfined phase of Yang-Mills
theories~\cite{YM1}, where, at high-$T$, the gluon mass can be
parametrized by $T/\ln T$.  The bending of the thermal masses can be
better appreciated in {}Fig.~\ref{fig4}, which shows that the changing
of behavior occurs at rather low temperatures.

\begin{figure}[htb!]
\includegraphics[scale=0.68]{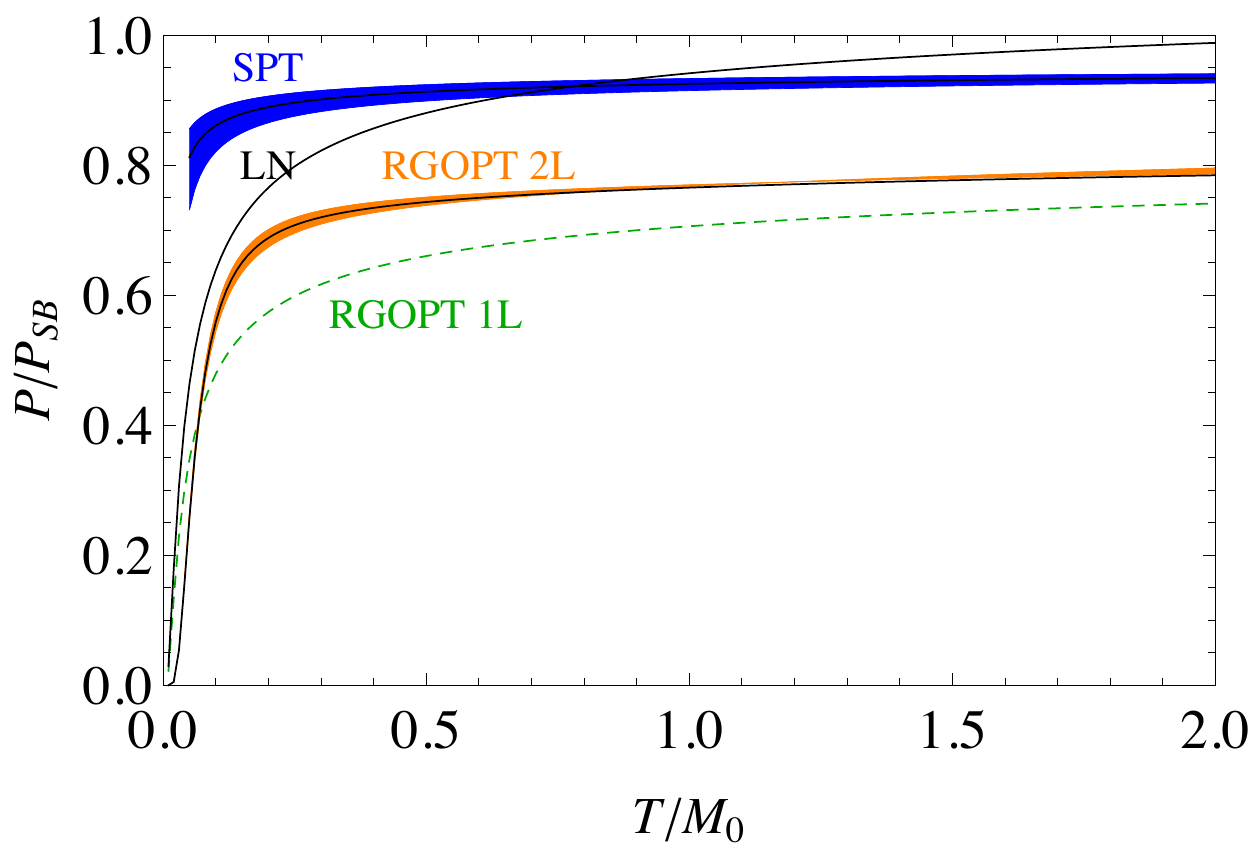}
\caption{\label{fig5} $P/P_{\rm SB}$ as function of the temperature
  $T$ (normalized by $M_0$) for $N=4$ and $g(M_0)=1 = g_{LN}(M_0)/2$,
  with scale variation $0.5\leq \alpha \leq 2$.  Within the two-loop
  RGOPT and SPT, the shaded bands  have the lower edge for
  $\alpha=0.5$ and the upper edge for $\alpha=2$. The thin line inside
  the shaded bands  is for $\alpha=1$.}
\end{figure}

In {}Fig.~\ref{fig5} we show the (subtracted) pressure, $P = P(T)-P(0)$,
normalized by $P_{\rm SB}$, for the scale variations $M=\alpha M_0$,
$0.5 \leq \alpha \leq 2$ and $N=4$.  It illustrates how the one-loop
RGOPT pressure is exactly scale invariant, while the two-loop result
displays a (small) residual scale dependence for the construction of the method (see \cite{ournlsm} for details). The RGOPT pressure itself exhibits a substantially smaller scale
dependence than the corresponding SPT approximation, at moderate and low $T/M_0$ values, 
as can be seen on {}Fig.~(\ref{fig5}). On \cite{ournlsm} we propose a temperature-dependent coupling $g(T)$ to minimize this moderate residual scale dependence within the two-loop results.

\begin{figure}[htb!]
\includegraphics[scale=0.67]{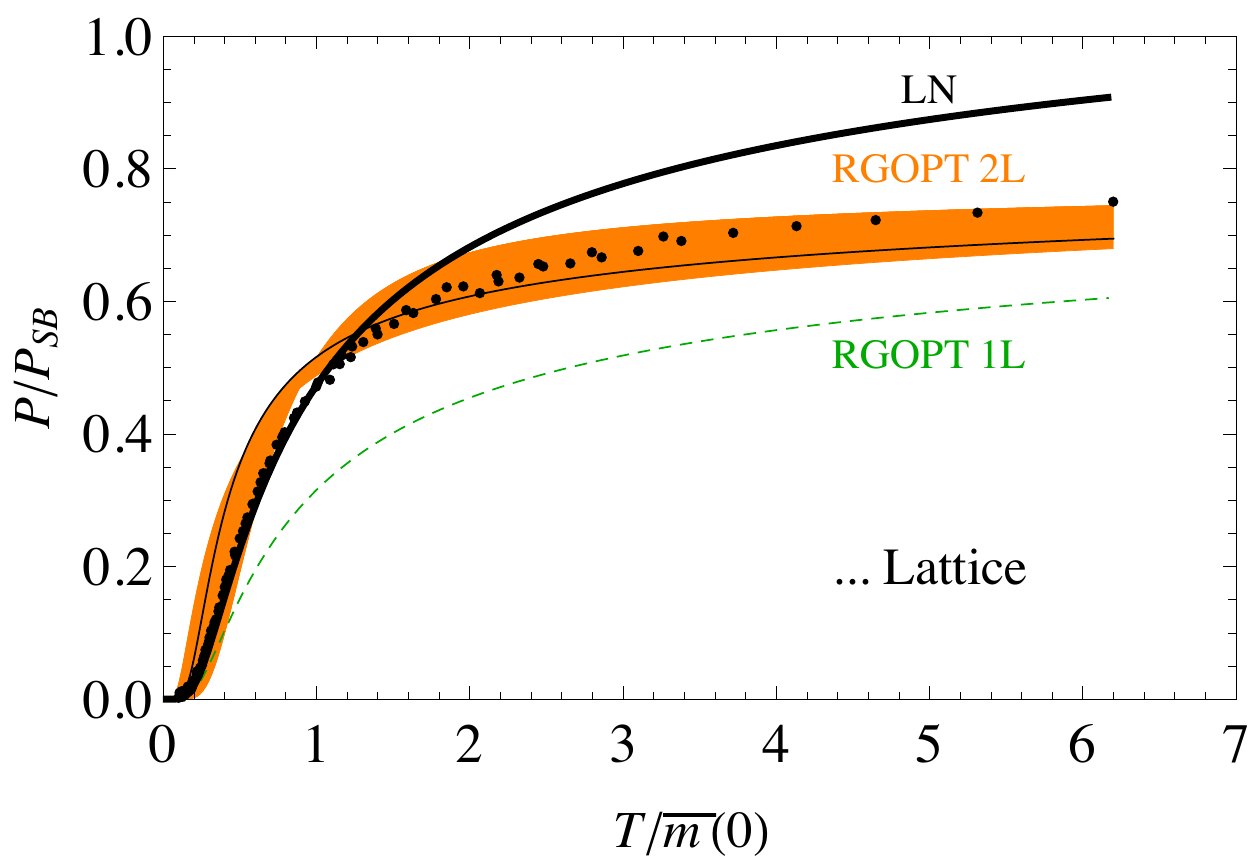}
\caption{\label{fig9} $P/P_{\rm SB}$ as a function of the
  temperature $T$ (normalized by the $T=0$ mass gap $\bar m(0)$) for $N=3$: LN, one- and
  two-loop RGOPT for  $g(M_0)=g_{LN}(M_0)=2\pi$ and scale variation $1/2 \leq
  \alpha \leq 2$, when using RG optimized running, compared with lattice simulations (taken from \cite{giacosa}).  NB lattice data have been conveniently
  rescaled on vertical axis from $P/T^2$ in \cite{giacosa} to
  $P/P_{SB}$ (i.e., for $N=3$ this corresponds to a scaling factor of
  $\pi/3$).}
\end{figure}

We will now compare the RGOPT results with lattice simulation ones. To the best of our knowledge, recently the only
available lattice thermodynamics simulation of the NLSM is the one of \cite{giacosa}, which was performed for
$N=3$.  
    To complete this comparison, we need  a priori to fix an
appropriate coupling value at some scale $M_0$, recalling that the simulation in \cite{giacosa} was performed at relatively strong lattice
coupling values. 
  In the RGOPT framework, similarly to the LN approximation,  as we have explained
  the constant vacuum energy piece $m^2/g$ (footprint of a $\sigma$ field term), plays a crucial
  role in obtaining a mass gap with these expected features of the low-$T$ nonperturbative NLSM properties. 
  While at asymptotically high-$T$ one reaches the free theory $g\to 0$ limit of the NLSM model, thus
  describing a gas of $N-1$ non-interacting pions, while the non-kinetic $m^2/g$ contribution becomes negligible.   
  The RGOPT two-loop results roughly exhibit this overall nonperturbative behavior from low- to high-$T$ regime (although not
  perfectly at very low temperatures). In
{}Fig.~\ref{fig9} we thus compare the one- and two-loop RGOPT
and the LN pressure for $g(M_0)=2\pi$  with the lattice data for
$N=3$,  as function of the temperature, now normalized  by 
the $T=0$ mass gap $\bar m(0)$, consistently with the lattice results normalization\cite{giacosa}. A clear explanation concerning some peculiarities about the $N$ choice can be found on \cite{ournlsm}.

It is also of interest to investigate the behavior of some other
thermodynamical quantities evaluated in the RGOPT scheme and how they
compare with the same quantities evaluated in the SPT and LN
approximations.  {}For example, the  interaction measure $\Delta=
({\cal E}- P)/T^2 \equiv T\partial_T[P(T)/T^2]$, 
which is the trace of the energy-momentum tensor
normalized by $T^2$. The  interaction measure  
can be readily obtained from the pressure
by using the definitions for the entropy density, 

\be 
S= \frac{d}{d\,T} P(\bar m(g,T,M),T,M),
\label{Sdef}
\ee 
and for   the energy density, ${\cal  E}= -P +S\, T$.

\begin{figure}[htb!]
\includegraphics[scale=0.67]{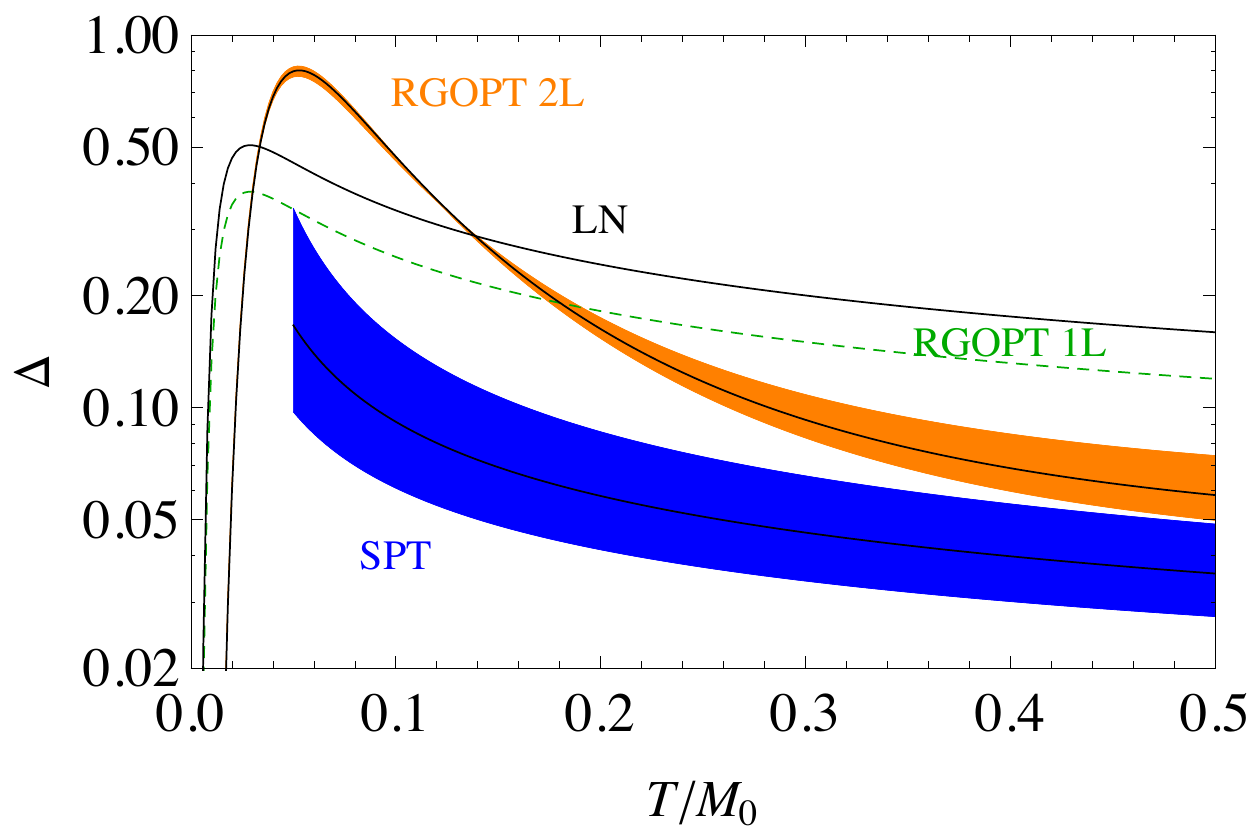}
\caption{\label{fig10} The interaction measure $\Delta$ as
  a function of the temperature $T$ (normalized by $M_0$)  for $N=4$
  and $g(M_0)=1 = g_{LN}(M_0)/2$, with scale variation  $0.5\leq
  \alpha \leq 2$, using the standard two-loop running coupling given
  by Eq.~\ref{run2}).  Within the two-loop RGOPT and SPT, the shaded
  bands  have the lower edge for $\alpha=0.5$ and the upper edge for
  $\alpha=2$. The thin line inside the shaded bands  is for
  $\alpha=1$. (Logarithmic scale is used).  }
\end{figure}

In  {}Fig.~\ref{fig10} we show the dependence of the interaction
measure as a function of the temperature in the one- and two-loop
RGOPT, two-loop SPT, and LN cases, for the same  choice of $N=4$ and
$g=1=g_{LN}/2$, as in the previous plots.

We notice from the RGOPT results shown in {}Fig.~\ref{fig10} how the
inflection before the peak  of $\Delta$ occurs approximately at the
temperature value where $m(T)$ bends (see {}Fig.~\ref {fig4}), which
is an interesting feature if one recalls that in QCD the inflection
occurs at $T_c$.

\section{Conclusions }
\label{sec7}

We have applied the recently developed RGOPT nonperturbative framework
to investigate thermodynamical properties of the asymptotically free
$O(N)$ NLSM in two dimensions, and illustrate results for $N=3$ and $N=4$. Our application shows
how simple perturbative results can acquire a robust nonperturbative predictive power 
by combining  renormalization group properties with a variational criterion used to fix the (arbitrary)
``quasi-particle" RGOPT mass. 

Our application shows
how simple perturbative results can acquire a robust nonperturbative predictive power 
by combining  renormalization group properties with a variational criterion used to fix the (arbitrary)
``quasi-particle" RGOPT mass. A non-trivial scale invariant result was obtained by
considering the lowest order contribution to the pressure and the NLO (two-loop) order RGOPT results display a very
mild residual scale dependence when compared to the standard SPT/OPT
results. We also obtain a reasonable agreement of the RGOPT pressure with known lattice results for $N=3$. The NLSM thermodynamical observables obtained from two-loop RGOPT display  a physical behavior that is more in line
with LQCD predictions for pure  Yang-Mills four-dimensional theories, as compared with the two-loop order SPT. The one- and two-loop RGOPT interaction measure $\Delta$ exhibit some characteristic nonperturbative
features somewhat similar to the QCD interaction measure.  Finally it would be of much interest to compare our NLSM thermodynamical results 
 with other lattice simulation results for other $N$ values,
  but unfortunately to our knowledge no such simulations at finite temperature are available up to now for $N>3$.

\section*{Acknowledgments}

GNF thanks CNPq for a PhD scholarship, and the Laboratoire Charles Coulomb
in Montpellier for the hospitality.


\end{document}